\documentclass{ws-ijmpa}

\begin{document}

\markboth{A.E. Dorokhov}
{Vector and Axial-Vector correlators in a instanton-like quark model}

%
\catchline{}{}{}{}{}
%

\title{VECTOR AND AXIAL-VECTOR CORRELATORS IN\\ A INSTANTON-LIKE QUARK MODEL}
\author{\footnotesize A.E. Dorokhov}

\address{Bogoliubov Laboratory of Theoretical Physics, Joint Institute for Nuclear
Research,\\ 141980, Dubna, Moscow Region, Russia}

\maketitle

\pub{Received 16 July 2004}{}

\begin{abstract}
The behavior of the vector Adler function at spacelike momenta is
studied in the framework of a covariant chiral quark model with instanton-like
quark-quark interaction. This function describes the transition between the
high energy asymptotically free region of almost massless current quarks to
the low energy hadronized regime with massive constituent quarks. The model
reproduces the Adler function and $V-A$ correlator extracted from the ALEPH
and OPAL data on hadronic $\tau$ lepton decays, transformed into the Euclidean
domain via dispersion relations. The leading order contribution from hadronic
part of the photon vacuum polarization to the anomalous magnetic moment of the
muon, $a_{\mu}^{\mathrm{hvp}\left(  1\right)}  $, is estimated.
\end{abstract}

In the chiral limit, where the masses of $u$, $d$, $s$ light quarks are set to
zero, the vector ($V$) and axial-vector ($A$) current-current correlation
functions in the momentum space (with $-q^{2}\equiv Q^{2}\geq0$) are defined
as
\begin{eqnarray}
\Pi_{\mu\nu}^{J,ab}(q) &  =i\int d^{4}x~e^{iqx}\Pi_{\mu\nu}^{J,ab}%
(x)=\,\left(  q_{\mu}q_{\nu}-g_{\mu\nu}q^{2}\right)  \Pi_{J}(Q^{2})\delta
^{ab},\label{PA}\\
\qquad\Pi_{\mu\nu}^{J,ab}(x) &  =\langle0\left\vert T\left\{  J_{\mu}%
^{a}(x)J_{\nu}^{b}(0)^{\dagger}\right\}  \right\vert 0\rangle,\nonumber
\end{eqnarray}
where the QCD $V$ and $A$ currents are
$
J_{\mu}^{a}=\overline{q}\gamma_{\mu}\frac{\lambda^{a}}{\sqrt{2}}q,\
J_{\mu}^{5a}=\overline{q}\gamma_{\mu}\gamma_{5}\frac{\lambda^{a}}{\sqrt{2}%
}q,\label{JAV}%
$
and $\lambda^{a}$ are Gell-Mann matrices $\left(  \mathrm{tr}\lambda
^{a}\lambda^{b}=2\delta^{ab}\right)  $. The momentum-space two-point
correlation functions obey (suitably subtracted) dispersion relations,
\begin{equation}
\Pi_{J}(Q^{2})=\int_{0}^{\infty}\frac{ds}{s+Q^{2}}\frac{1}{\pi}\mathrm{Im}%
\Pi_{J}(s),\label{Peuclid}%
\end{equation}
where the imaginary parts of the correlators determine the spectral functions
$
\rho_{J}(s)=4\pi\mathrm{\Im}\Pi_{J}(s+i0).\nonumber
$
Instead of the correlation function it is more convenient to work with the
Adler function defined as
\begin{equation}
D_{J}(Q^{2}){=-Q}^{2}\frac{d\Pi_{J}(Q^{2})}{dQ^{2}}{\,=}\frac{1}{4\pi^{2}}%
\int_{0}^{\infty}{dt}\frac{Q^{2}}{(t+Q^{2})^{2}}{\,\rho_{J}(t)\,.}%
\label{Adler}%
\end{equation}


Our goal is to obtain the vector current-current correlator and corresponding
Adler function in the low and intermediate regions of momenta
by using the effective model and then to estimate the leading
order hadron vacuum polarization correction to $a_{\mu}$.
In N$\chi$QM in the chiral limit the correlators have transverse character%
\begin{equation}
\Pi_{\mu\nu}^{J}\left(  Q^{2}\right)  =\left(  g_{\mu\nu}-\frac{q^{\mu}q^{\nu
}}{q^{2}}\right)  \Pi_{J}^{\mathrm{N\chi QM}}\left(  Q^{2}\right)
,\label{PVmn}%
\end{equation}
where the polarization functions are given by the sum of the dynamical quark
loop, the intermediate (axial-)vector mesons and the higher order mesonic
loops contributions
\begin{equation}
\Pi_{J}^{\mathrm{N\chi QM}}\left(  Q^{2}\right)  =\Pi_{J}^{Q\mathrm{Loop}%
}\left(  Q^{2}\right)  +\Pi_{J}^{\mathrm{mesons}}\left(  Q^{2}\right)
+\Pi_{J}^{\chi\mathrm{Loop}}\left(  Q^{2}\right)  ,\label{Pncqm}%
\end{equation}
where the mesons are generated via resummation of quark loop chains.


The dominant contribution to the vector current correlator is given by the
dynamical quark loop which was found in \cite{Dorokhov:2003kf,Dorokhov:2004ze} with the
result
\begin{eqnarray}
\Pi_{V}^{Q\mathrm{Loop}}\left(  Q^{2}\right)   &  =\frac{4N_{c}}{Q^{2}}%
\int\frac{d^{4}k}{\left(  2\pi\right)  ^{4}}\frac{1}{D_{+}D_{-}}\left\{
M_{+}M_{-}+\left[  k_{+}k_{-}-\frac{2}{3}k_{\perp}^{2}\right]  _{ren}\right.
\label{Ploop}\\
&  +\left.  \frac{4}{3}k_{\perp}^{2}\left[  \left(  M^{\left(  1\right)
}\left(  k_{+},k_{-}\right)  \right)  ^{2}\left(  k_{+}k_{-}-M_{+}%
M_{-}\right)  -\left(  M^{2}\left(  k_{+},k_{-}\right)  \right)  ^{\left(
1\right)  }\right]  \right\}  +\nonumber\\
&  +\frac{8N_{c}}{Q^{2}}\int\frac{d^{4}k}{\left(  2\pi\right)  ^{4}}%
\frac{M\left(  k\right)  }{D\left(  k\right)  }\left[  M^{\prime}\left(
k\right)  -\frac{4}{3}k_{\perp}^{2}M^{\left(  2\right)  }\left(
k,k+Q,k\right)  \right]  ,\nonumber
\end{eqnarray}
where the notations
$
k_{\pm}=k\pm Q/2, k_{\perp}^{2}=k_{+}k_{-}-\frac{\left(  k_{+}q\right)
\left(  k_{-}q\right)  }{q^{2}}, D\left(  k\right)  =k^{2}+M^{2}(k),
$
and $
M_{\pm}=M(k_{\pm}),\  D_{\pm}=D(k_{\pm})
$
are used. We also introduce the finite-difference derivatives defined for an
arbitrary function $F\left(  k\right)  $ as
\begin{equation}
F^{(1)}(k,k^{\prime})=\frac{F(k^{\prime})-F(k)}{k^{\prime2}-k^{2}},\qquad
F^{(2)}\left(  k,k^{\prime},k^{\prime\prime}\right)  =\frac{F^{(1)}%
(k,k^{\prime\prime})-F^{(1)}(k,k^{\prime})}{k^{\prime\prime2}-k^{\prime2}}.
\label{FDD}%
\end{equation}

\begin{figure}[h]
\hspace*{-1cm} \begin{minipage}{7cm}
\vspace*{0.5cm} \epsfxsize=6cm \epsfysize=5cm \centerline{\epsfbox
{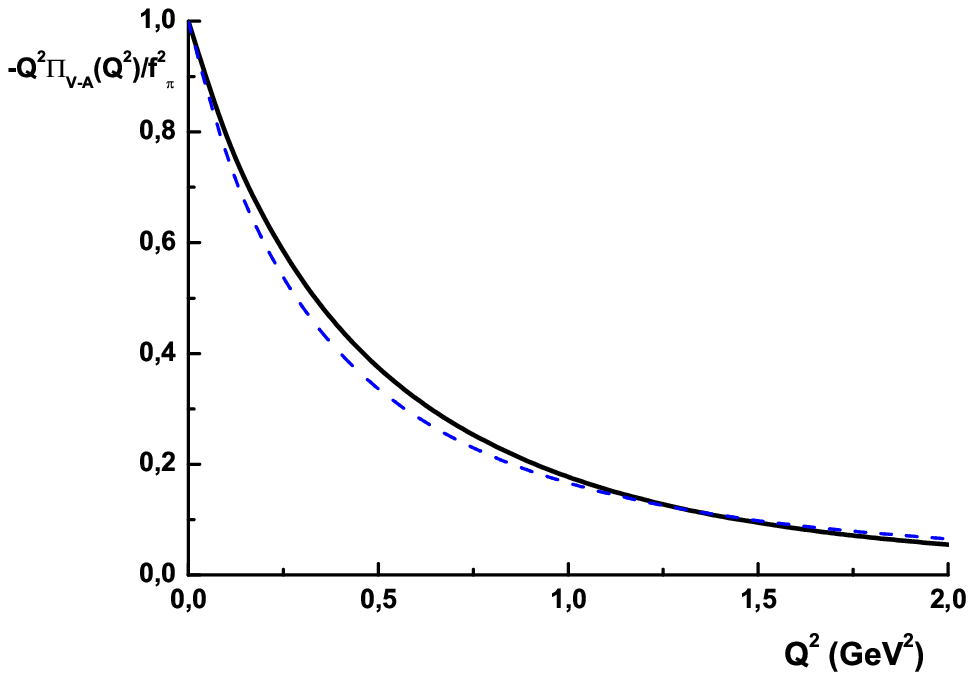}}
\caption[dummy0]{Normalized $V-A$
correlation function constructed in the N$\chi$QM (solid line) and
reconstructed from the ALEPH experimental spectral function (dashed line).}%
\label{Fv-a}
\end{minipage}\hspace*{0.5cm} \begin{minipage}{7cm}
\vspace*{0.5cm} \epsfxsize=6cm \epsfysize=5cm \centerline{\epsfbox{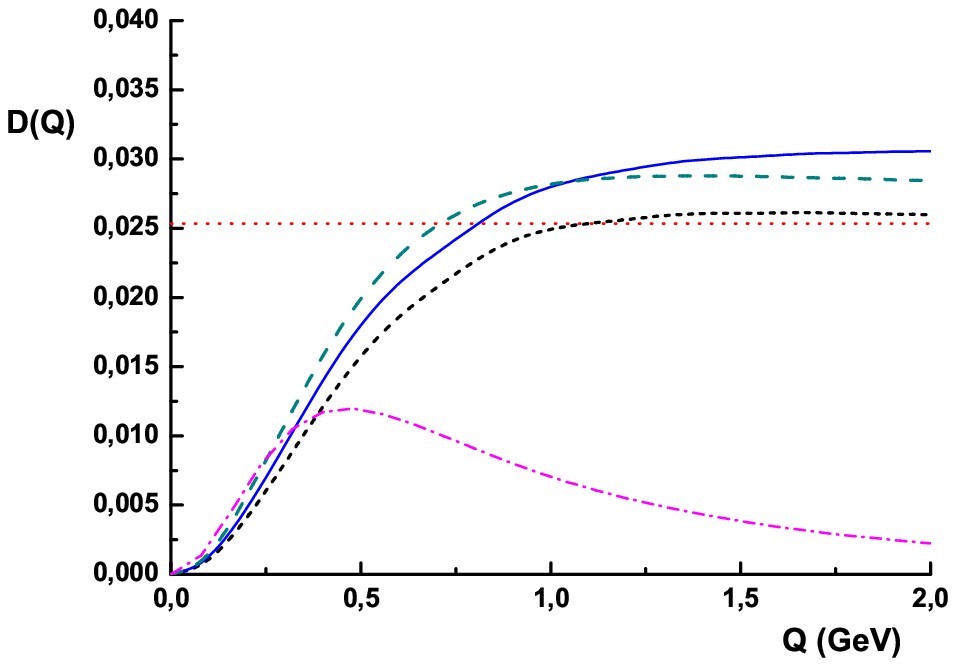}}
\caption[dummy0]{The Adler function from
the N$\chi$QM contributions: dynamical quark loop (short dashed), quark +
chiral loops + vector mesons (full line) versus the ALEPH data (dashed). The
dash-dotted line is the prediction of the constituent quark model (ENJL) and
the dotted line is the asymptotic freedom prediction, $1/4\pi^{2}$.}\label{AdlerV}
\end{minipage}\end{figure}

The difference of the V and A correlators is free from any perturbative corrections
for massless quarks and very sensitive to the spontaneous breaking of chiral symmetry.
The model calculations of the chirality flip combination provides
\begin{eqnarray}
\Pi_{V-A}^{\mathrm{Loop}}\left(  Q^{2}\right)   &  =-\frac{4N_{c}}{Q^{2}}%
\int\frac{d^{4}k}{\left(  2\pi\right)  ^{4}}\frac{1}{D_{+}D_{-}}\left\{
M_{+}M_{-}+\frac{4}{3}k_{\perp}^{2}\left[  -\sqrt{M_{+}M_{-}}M^{\left(
1\right)  }\left(  k_{+},k_{-}\right)  +\right.  \right.  \label{VmAmodel}\\
&  \left.  \left.  +\left(  \sqrt{M}^{\left(  1\right)  }\left(  k_{+}%
,k_{-}\right)  \right)  ^{2}\left(  \sqrt{M_{+}}k_{+}+\sqrt{M_{-}}%
k_{-}\right)  ^{2}\right]  \right\}  .\nonumber
\end{eqnarray}
and the results extracted from ALEPH experiment \cite{ALEPH2}
is shown in Fig. \ref{Fv-a}. Oppositely each correlator separately is dominated by perturbative
massless quark loop diagram in the high momenta region. In the model calculations this result
is reproduced because the dynamical quark mass, $M(k)$ at large virtualities $k^2$ vanishes
in the chiral limit.
The results of the model calculation of the vector Adler function and its experimental conterpart
from ALEPH and perturbative QCD asymptotics (see details in \cite{Dorokhov:2004ze}) are presented in
Fig.  \ref{AdlerV}

Determination of the Adler function allows us to estimate the leading order hadronic
vacuum contribution to the muon anomalous magnetic moment by using the integral representation
\begin{equation}
a_{\mu}^{\mathrm{hvp}\left(  1\right)  }=\frac{4}{3}\alpha^{2}\int_{0}%
^{1}dx\frac{\left(  1-x\right)  \left(  2-x\right)  }{x}D_{V}\left(
\frac{x^{2}}{1-x}m_{\mu}^{2}\right)  ,\label{aAd}%
\end{equation}
where the charge factor $\sum Q_{i}^{2}=2/3$, $i=u,d,s,$ is taken into
account. One gets the model estimate
\begin{equation}
a_{\mu}^{\mathrm{hvp}\left(  1\right)  }=6.53\cdot10^{-8}\label{ammALEPH}%
\end{equation}
which is in a reasonable agreement with the precise phenomenological numbers,
found from precise determination of the low energy tail of the total $e^{+}
e^{-}\rightarrow$ hadrons and $\tau$ lepton decays cross-sections\cite{ynd}
\begin{equation}
a_{\mu}^{\mathrm{hvp}\left(  1\right)  }=\left\{
\begin{array}
[c]{c}%
(6.935\pm0.09)\cdot10^{-8}\qquad e^{+}e^{-},\\
(7.018\pm0.09)\cdot10^{-8}\qquad\tau.
\end{array}
\right.
\end{equation}
As by product we estimate also the anomalous magnetic moment of the $\tau$ lepton as
\begin{equation}
a_{\tau}^{\mathrm{hvp}\left(  1\right)  }=3.06\cdot10^{-6}.\label{ammALEPH}%
\end{equation}

We plan to apply the nonlocal chiral quark model to get
reliable estimates of the higher order hadronic vacuum polarization to the
anomalous magnetic moments of leptons, where there are no precise phenomenological
determinations of these contributions.

Ther author thanks for partial support from RFBR Grant no. 02-02-16194.

\end{document}